# Molecular alignment echoes probing collision-induced rotational-speed changes

J.-M. Hartmann[1#], J. Ma[2,3], T. Delahaye[1], F. Billard[2], E. Hertz[2],
J. Wu[3,4], B. Lavorel[2], C. Boulet[5], O. Faucher[2*]

[1]Laboratoire de Météorologie Dynamique/IPSL, CNRS, Ecole polytechnique, Institut polytechnique de Paris, Sorbonne Université, Ecole Normale Supérieure, PSL Research University, F-91128 Palaiseau, France
[#]jean-michel.hartmann@lmd.polytechnique.fr

[2]Laboratoire Interdisciplinaire CARNOT de Bourgogne,
UMR 6303 CNRS-Université de Bourgogne Franche-Comté, BP 47870, 21078 Dijon, France
[*]olivier.faucher@u-bourgogne.fr

[3]State Key Laboratory of Precision Spectroscopy, East China Normal University, Shanghai 200062, China

[4]Collaborative Innovation Center of Extreme Optics, Shanxi University, Taiyuan, Shanxi 030006, China

[5]Université Paris-Saclay, CNRS, Institut des Sciences Moléculaires d'Orsay, 91405, Orsay, France

**Abstract:** We show that the decays with pressure of the rotational alignment echoes induced in $N_2O$-He gas mixtures by two ultrashort laser pulses with various delays show detailed information about collision-induced changes of the rotational speed of the molecules. Measurements and classical calculations consistently demonstrate that collisions reduce the echo amplitude all the more efficiently when the echo appears late. We quantitatively explain this behavior by the filamentation of the classical rotational phase space induced by the first pulse and the narrowing of the filaments with time. The above mentioned variation of the echo decay then reflects the ability of collisions to change the molecular rotation speed by various amounts, enabling refined tests of models for the dissipation induced by intermolecular forces. We also demonstrate that the collision-induced changes of the rotational speed within the filaments are the classical equivalents of the nonsecular transfers among quantum coherences, thus evidencing the correspondence between the classical and quantum worlds.

## I. INTRODUCTION

In gas media, intermolecular collisions modify the rotational and translational motions of molecules. These processes are of fundamental interest and must be properly modeled for practical applications. Indeed, in gases at equilibrium, molecule-molecule and -atom interactions largely govern the shape of molecular spectra [1] and the energy and mass transports [2,3], for instance. When the system has been perturbed by an external interaction, the return to equilibrium often follows channels enabled by collisions [2].

Much previous research has studied collisional effects in the frequency domain through their consequences on the shape of light-absorption and -scattering spectra [1,4]. The influences of intermolecular forces in gases have also been investigated in the time domain through the decay of the molecular-axis alignment/orientation induced by an electromagnetic pulse (e.g. Refs. [5-12]) and, recently, by using the echo following the excitation of the system by two laser kicks [13,14]. However, these studies provide averaged information with little detail on intermolecular interactions. For instance, the pressure broadening of anisotropic Raman lines or the decay of alignment revivals (which are, through a Fourier transform, equivalent [11]) only tell how fast collisions change the rotational motion, with no information on the respective contributions of dephasing, changing of the rotational speed or of the angular momentum orientation. The "transient" and "permanent" components of alignment signals provide more insights as they tell at which (different) time scales relaxation affects the modulus and orientation of the rotational angular momentum [8,10,12,15]. However, again, no information is provided on how efficiently do collisions modify the rotational energy by a given amount. The way to overcome these limitations was, up to now, to carry joined time and spectral domains (double resonance) experiments [16,17]. One can, for instance, depopulate a rotational level by a resonant pulse and then probe the evolutions of other levels populations using the time and spectral dependences of the absorption spectrum. Such measurements can provide state-to-state relaxation rates, but they are complicated to carry. Furthermore, disentangling the contributions of the various collisional channels requires the use of a model.

This paper gives experimental and theoretical demonstrations that alignment echoes enable detailed investigations of the collision-induced changes of molecular rotation. These echoes in the alignment of molecular axes after nonadiabatic excitations by two successive nonresonant and linearly-polarized laser pulses were shown in low pressure $CO_2$ [18]. Some of the molecules are first aligned by a short and intense pulse ($P_1$) thanks to the anisotropy of the molecular polarizability. This results in a peak in the alignment factor $<\cos^2(\theta)>(t)$, where $\theta$ is the angle between the molecule axis and the laser polarization, which vanishes quickly due to the spread of the angular velocities. A second pulse ($P_2$), applied $\tau_{12}$ later, induces a rephasing creating an echo in the alignment factor at $t = 2\tau_{12}$ after $P_1$ [e.g. Fig. 1(a)]. Further investigations, still carried at pressures for which collisional effects are negligible at the investigated time scales, revealed the existence of fractional, imaginary, and rotated echoes [19-21]. The collisional dissipation of echoes was studied later [13,14], by varying the delay $\tau_{12}$ for fixed gas densities. A pressure-induced decay time constant was obtained from the decrease, with increasing $\tau_{12}$, of the echo amplitude. Reference [14] demonstrated the advantage of this approach under high-pressure for which the alignment revivals have vanished and cannot be used to probe the collisional dynamics. However, such experiments provide a single time constant reflecting the influence of all types of rotation-changing collisions throughout the time window of observation. The echoes were revisited very recently from a much richer point of view, by studying the decay of their amplitudes with increasing pressure for fixed delays $\tau_{12}$ [22]. This provided the evolution, with $\tau_{12}$, of the echo dissipation, showing that the efficiency of collisions in damping the echo significantly varies with time at the early stage of the relaxation process. Associated quantum theoretical calculations evidenced the breakdown of the widely used "secular approximation" at short times.

We here reanalyze the experimental results of Ref. [22] with a completely different approach, using a classical instead of a quantum model. This reveals that the variation of the echo pressure-induced decay with its time of appearance reflects the ability of collisions to change the rotational speed by various amounts. We also show that rotational alignment echoes are some kind of bridge between the quantum and classical worlds, as also recently observed in an isolated vibrating molecule [23]. For this, we demonstrate the tight correspondence between the progressive filamentation of the angular speed in the classical rotational phase space and the time evolution of the quantum coherences created by the first laser pulse. We also show that the validity of the secular approximation, quantum mechanically driven by the dephasing between the coherences is, in the classical world, conditioned by the width of the filaments.

## II. EXPERIMENTS

The experimental set-up, the procedures used for the measurements and their analysis, and the results used below have been presented in Ref. [22]. Two linearly polarized, intense, and non resonant femtosecond laser pulses were applied, separated by various delays $\tau_{12}$, and the alignment factor was recorded for $N_2O(4\%)+He(96\%)$ gas mixtures at room temperature and several pressures. For this, a third (weak) pulse was used, with various delays with respect to the pumps, that probed the transient birefringence resulting from the anisotropic distribution of molecular orientations. The modification of its polarization was analyzed by a highly sensitive balanced detection delivering a signal proportional to $<\cos^2(\theta)-1/3>(t)$. As exemplified by the calculated signal of Fig. 1(a) (and by the measurement in Fig. 2(a) of [22]), the first pulse, at $t=t_1$, results, thanks to the molecular polarizability anisotropy, in an alignment peak that quickly disappears due to the dephasing of the quantum coherences that it has generated. The second pulse, at $t=t_1+\tau_{12}$, which also generates a quickly vanishing alignment, somehow reverses the course of time and induces a progressive rephasing of the coherences (see Appendix) leading to the formation of an alignment echo at $t=t_1+2\tau_{12}$. The variation of the amplitude of this echo with the gas density contains information on the loss of coherences induced by inter-molecular collisions [22]. In order to quantify this process, after normalization by the gas density $d$ in order to remove the proportionality of the signal to the number of molecules in the volume excited by the lasers, the amplitude $S(d,\tau_{12})$ (see Fig. 1b) of each measured echo was determined. The values obtained for several densities were then fitted by $S(d,\tau_{12}) = A(\tau_{12})\exp[-d/d_0(\tau_{12})]$ and a density-normalized time constant (in ps.amagat, 1 amagat corresponding to $2.69 \times 10^{25}$ molecules/m$^3$) was then defined by $\tau_E(\tau_{12}) = 2\tau_{12} d_0(\tau_{12})$, which accounts for the fact that the echo appears at $t = 2\tau_{12}$ after the time origin defined by P$_1$.

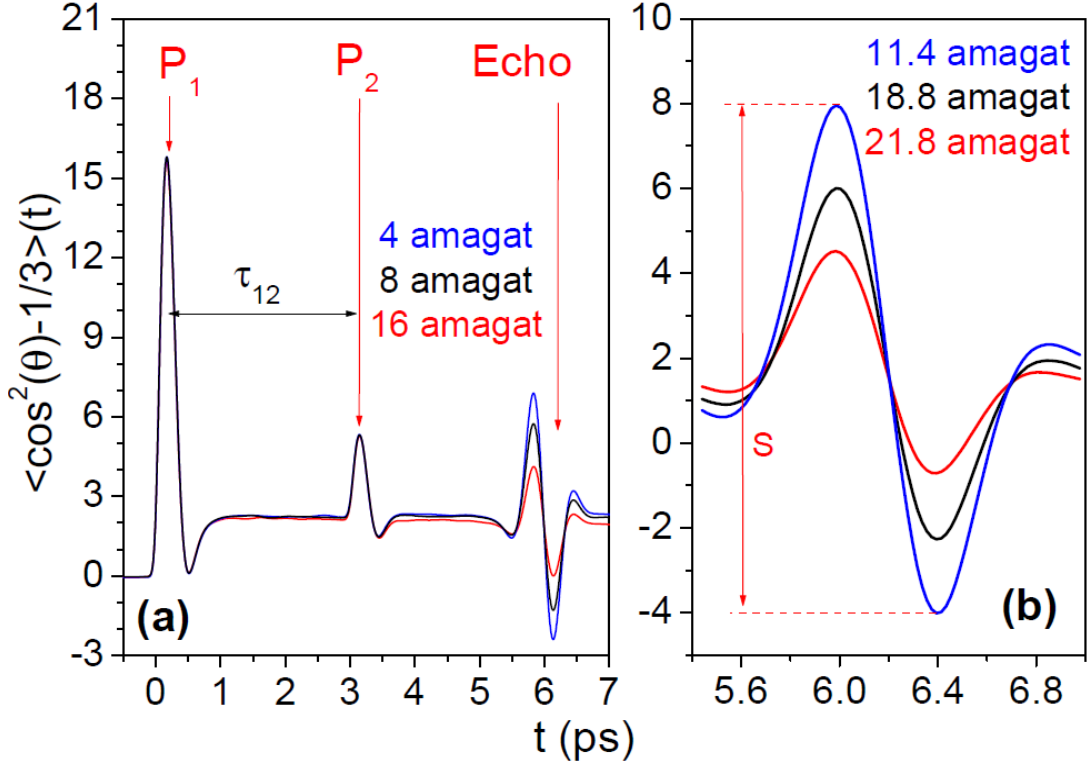

**Fig. 1:** Alignment factors for N$_2$O diluted in He at various densities. (a) Computed values for $\tau_{12}$ =3.0 ps. (b) Measured values for $\tau_{12}$ =3.1 ps.

### III. CALCULATIONS

Classical Molecular Dynamics Simulations (CMDS) were carried for N$_2$O diluted in He gas, as described in Refs. [14,15]. Many molecules and atoms were treated using periodic boundary conditions, nearest-neighbors' spheres, and the Verlet algorithm [24]. After proper initializations (Boltzmanian energies, random orientation of the vectors, molecules/atoms not to close to each other [15]) of the center of mass position $\vec{q}_m(t=0)$ and velocity $\vec{v}_m(t=0)$, and of the axis orientation $\vec{u}_m(t=0)$ and angular speed $\vec{\omega}_m(t=0)$ of each particle $m$, these quantities were propagated in time using the equations of *classical* mechanics. The force $\vec{F}_m(t)$ and torque $\vec{\tau}_m(t)$ applied to each molecule (or atom) by its neighbors and the laser pulses were computed, as explained in [15], from knowledge of the positions and orientations of all particles. $\vec{q}_m(t+dt)$, $\vec{v}_m(t+dt)$, $\vec{u}_m(t+dt)$, and $\vec{\omega}_m(t+dt)$ were then obtained from $\vec{F}_m(t)$, $\vec{\tau}_m(t)$, $\vec{q}_m(t)$, $\vec{v}_m(t)$, $\vec{u}_m(t)$, and $\vec{\omega}_m(t)$. In order to simulate molecules infinitely diluted in He while keeping computer time reasonable, 50% N$_2$O + 50% He mixtures were treated by only taking N$_2$O-He forces into account with the accurate potential of Ref. [25]. The laser pulses characteristics were set to those of the experiments (linearly polarized along the same direction, Gaussian time-envelop with a 100 fs full width at half maximum (FWHM) and peak intensities around 20 TW/cm$^2$) and the N$_2$O anisotropic polarizability $\Delta\alpha$ =19.8 $a_0^3$ [26] was used. The molecules were treated as rigid rotors, and the requantization procedure introduced in Ref. [15] was eventually applied, in which the rotational angular speed is changed, at properly chosen times, to match the closest quantum value. These requantized CMDS (rCMDS) and the purely classical CMDS enabled to calculate the alignment factor $<\cos^2\theta(t)>$ for several densities [Fig. 1(a)]. The time constant $\tau_E(\tau_{12})$ of the echo decay was then determined from the predicted alignments as done with the experimental ones (Sec. II and Ref. [22]). Note that the merits of (r)CMDS have been demonstrated for various laser-induced alignment features in different gases [7,9,14,19,20,27,28].

## IV. RESULTS

The experimental and predicted values of $\tau_E(\tau_{12})$ are displayed in Fig. 2. As pointed out previously [22], the measured echo decay, which is relatively slow at short delays, becomes faster as the delay increases before a plateau is reached around $2\tau_{12} \approx 12$ ps with $\tau_E(2\tau_{12} \geq 12 \text{ ps}) \approx 70$ ps.amagat. This is well reproduced by the rCMDS despite a predicted plateau reached a little later (around $2\tau_{12} \approx 20$ ps) and slightly lower [with $\tau_E(2\tau_{12} \geq 20 \text{ ps}) \approx 63$ ps.amagat] than in the experiments. Concerning the CMDS, they also lead to very satisfactory predictions before about 10 ps but do not predict any plateau. This, which is discussed later, comes from the fact they enable filament widths (and rotational speed changes) to become smaller than the quantum limit $\hbar/I$. However, note that this quantum limit is reached when the time approaches the revival time for N$_2$O (20 ps, see Appendix A), which makes sense.

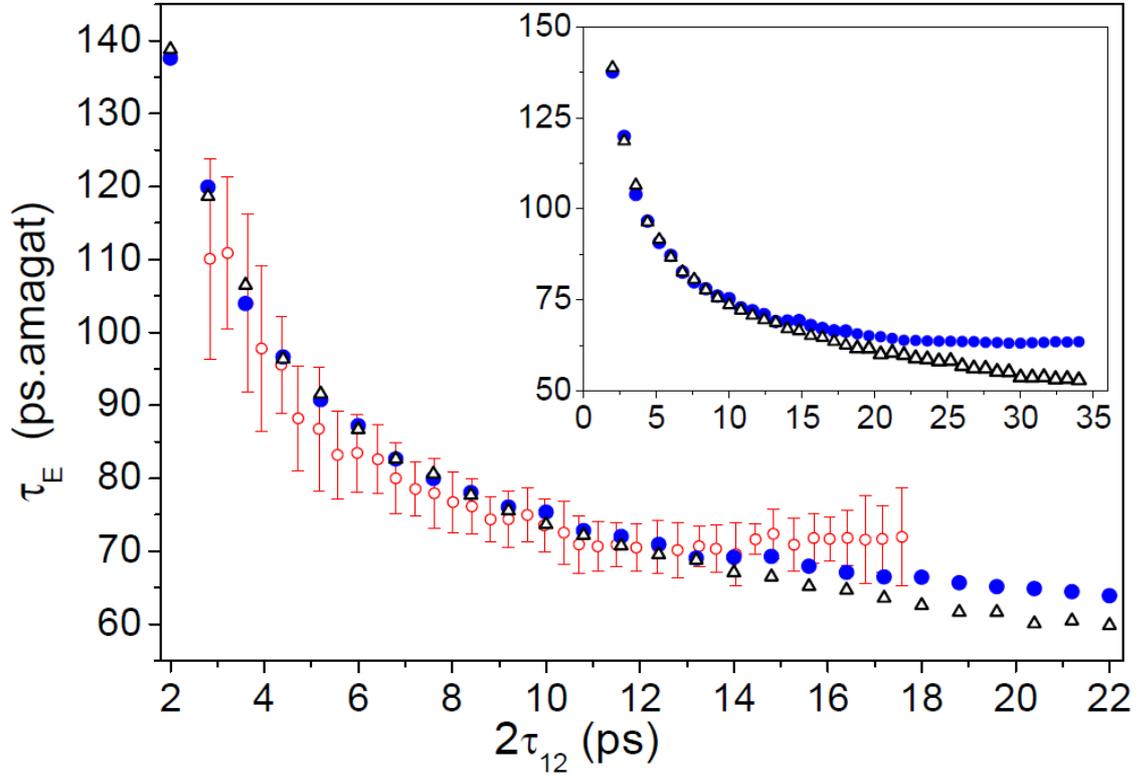

**Fig. 2:** Density-normalized decay time constants $\tau_E(\tau_{12})$ of the echoes for N$_2$O diluted in He at 295 K obtained from measured (red open circles, from [22]), CMDS- (black open triangles) and rCMDS-computed (blue full circles) alignment factors for various densities and fixed delays. The insert shows the predicted values over a broader time interval.

## V. DISCUSSION
### V.A Phase-space filamentation, echoes and rotational speed changes

Recall that the alignment echo created at $t=2\tau_{12}$ by laser kicks at $t=0$ and $t=\tau_{12}$ is a classical phenomenon [18], contrary to the revivals (see detailed demonstration in the Appendix where several experimental characteristics of the revivals and echoes are analytically derived). This echo results from the fact that the first pulse changes the molecule rotational speed according to the angle $\theta_0$ between its axis and the laser polarization at the time of the excitation, with the largest effect for $\theta_0$=45° and no response to the excitation for $\theta_0$=0° and $\theta_0$=90°. This induces [18] a progressive filamentation of the ($\omega,\theta$) rotational phase space which is explained using a two-dimensional 2D model in Refs. [18,19], and shown in Fig. 3 and in the Supplementary Material [29] (both obtained from 3D CMDS for $N_2O$ gas under collision-free conditions). The second pulse, at $t=\tau_{12}$ then somehow acts on the filaments in a way similar to the influence of the first pulse on the initial equilibrium distribution, leading to a folding of the filaments which bunch up near $\theta$=0 at $t=2\tau_{12}$ as can be seen in Fig. 3 and in [18,20],. In Fig. 3, the echo appears clearly in the alignment factor displayed in the right hand-side panel, with a maximum of the alignment at $t=t^+$=5.82 ps followed by a minimum at $t=t^-$=6.14 ps. The maximum results from the fact that, in the phase space (see left-hand side panel), the various folded filaments associated with high probability densities (appearing in orange, yellow and green) are aligned (along the horizontal dashed line at $t$=5.82 ps) and collectively contribute, which was not the case before. Symmetrically, a little later, it is the folded filaments with a low probability density (in blue) that are aligned (along the horizontal dashed line at $t$=6.14 ps), in contrast with the situation before and after, leading to a minimum alignment factor. Equivalents of these features within a 2D analytical model can be found in Fig. 3 of Ref. [18]. Note that the extrema in the probability distribution of the molecular axes associated with the maximum and minimum of the alignment echo can also bee see in the Supplementary Material [29], around $\theta$=0° for $t^*$=1.95 and around $\theta$=90° for $t^*$=2.05, respectively. Recall that fractional echoes also appear after the second pulse, at rational fractions of the delay (e.g. at $t=4\tau_{12}/3, 3\tau_{12}/2, …$) [19,20]. These can be seen in the supplementary movie where high-order transient symmetric structures appear in the angular distribution (e.g. around $t^*$=1.5), While these structures do not manifest in the alignment factor $<\cos^2(\theta)>(t)$ due to their high angular symmetry, they can be observed using other experimental probing techniques [19,21].

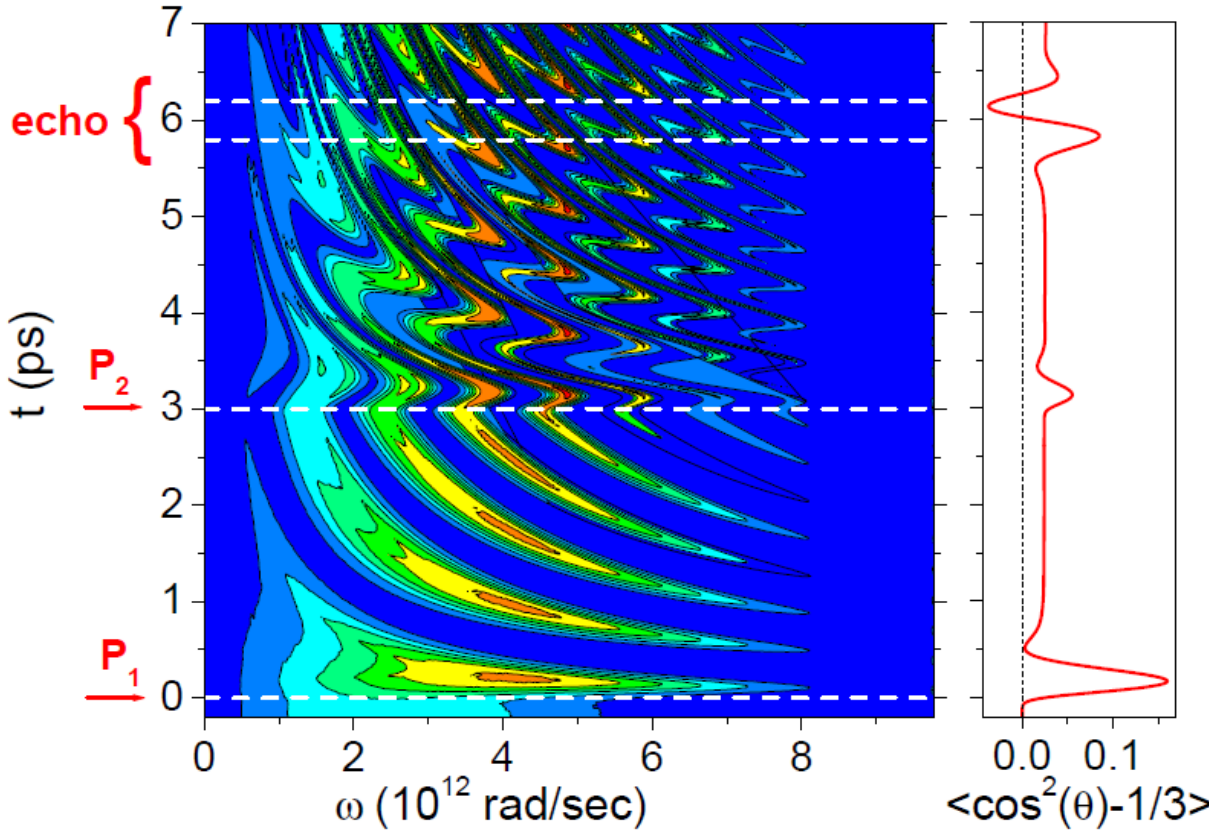

**Fig. 3:** Time-dependent CMDS-predictions for collision free $N_2O$ gas excited by laser pulses at $t$=0 and $t$=3 ps. Left panel: Probability density (from blue to red with increasing value) for the $N_2O$ molecules whose axis makes an angle $\theta$=0° with the lasers polarization to have an angular velocity $\omega$. Right panel: Alignment factor.

As shown by Fig. 4, the number $n(t)$ of filaments increases linearly with time after the pulse, while their mean width $\Delta\omega_{FWHM}(t)$ and the average distance $\Delta\omega_p(t)$ between their peaks reduce. Note that $\Delta\omega_p(t)$ very well follows the $\pi/t$ law deduced from a 2D model [18]). Note that, since the number of filaments is integer valued, its increase with time follows a stepped curve (looking like a stair-case) not represented in Fig. 4(b) where only the value at the center of each step is plotted (the same remark also applies to the mean distance between the filament). The time evolution of the filament width in Fig. 4(c) provides a first, but still only qualitative, explanation for the results in Fig. 2. Indeed, because it is the progressive removal of the molecules from "their" filament by collisions that induces the echo decay, it is obvious that this process is all the more rapid when the filaments are narrow.

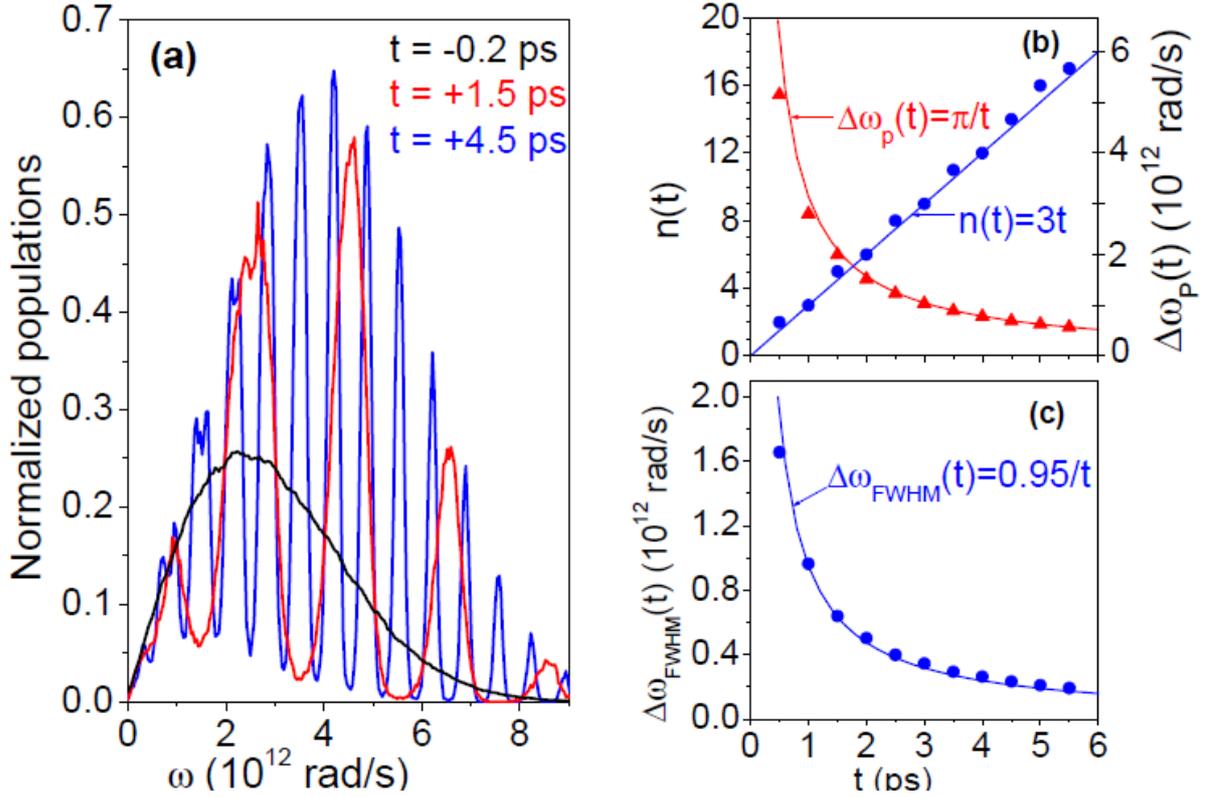

**Fig. 4:** CMDS results for collision-free $N_2O$ excited by a laser pulse centered at $t=0$. (a) Probability distribution (i.e. normalized number of molecules so that the integration over $\omega$ leads to a unit area) of angular velocities at different times for the molecules whose axis makes an angle of 0° with the laser polarization (the black line being the CMDS predicted Boltzmann distribution before the pulse); (b) number of filaments ($n$, blue full circles, left y-axis) and mean distance between successive filaments ($\Delta\omega_p$, red triangles, right y-axis) ; (c) Mean full width at half maximum $\Delta\omega_{FWHM}$ of the filaments.

In order to now be quantitative, we carried CMDS for $N_2O$ diluted in He gas at equilibrium (no pulse), and looked at the time evolution of the number of molecules whose angular speed was between $\omega_0 - \Delta\omega/2$ and $\omega_0 + \Delta\omega/2$ at $t=0$ and is still within this interval at $t$. The time constant $\tau_{Loss}(\Delta\omega)$ of the decay of this number is shown by the black line in Fig. 5 (obtained for the most probable angular speed $\omega_0 = 2.3\ 10^{12}$ rad/s ). As expected, $\tau_{Loss}(\Delta\omega)$ increases with $\Delta\omega$, since changing the rotational speed by a larger amount requires more time, because either a strong (and thus little probable) collision or several successive weak collisions are needed. To now cast the results of Fig. 2 into Fig. 5, we need to associate a representative filament width $\Delta\omega(\tau_{12})$ with each x-axis coordinate of Fig. 2. For this, and considering that the mean width of the filaments evolves with time [Fig. 4(c)], it seems reasonable to choose the value for the median time $t_M = [t(Echo) - t(P_1)]/2 = \tau_{12}$ between $P_1$ and the echo. We thus retained $\Delta\omega(\tau_{12}) = 2\Delta\omega_{FWHM}(\tau_{12})$, with $\Delta\omega(\tau_{12}) = 1.9/\tau_{12}$ [see Fig. 4(c)], where the factor 2 is introduced in order to include the full filament and not its top half only. Then plotting the measured values from Fig. 2 vs $\Delta\omega(\tau_{12})$ leads to the open circles in Fig. 5. As can be seen, despite its simplicity, the model based on the rate of removal of molecules from given rotational-speed intervals well agrees with the observed results. This quantitatively confirms the above given explanation of the results of Fig. 2 by the time evolution of the filaments widths, and the fact that molecules are more quickly removed from their filament when the latter is narrow.

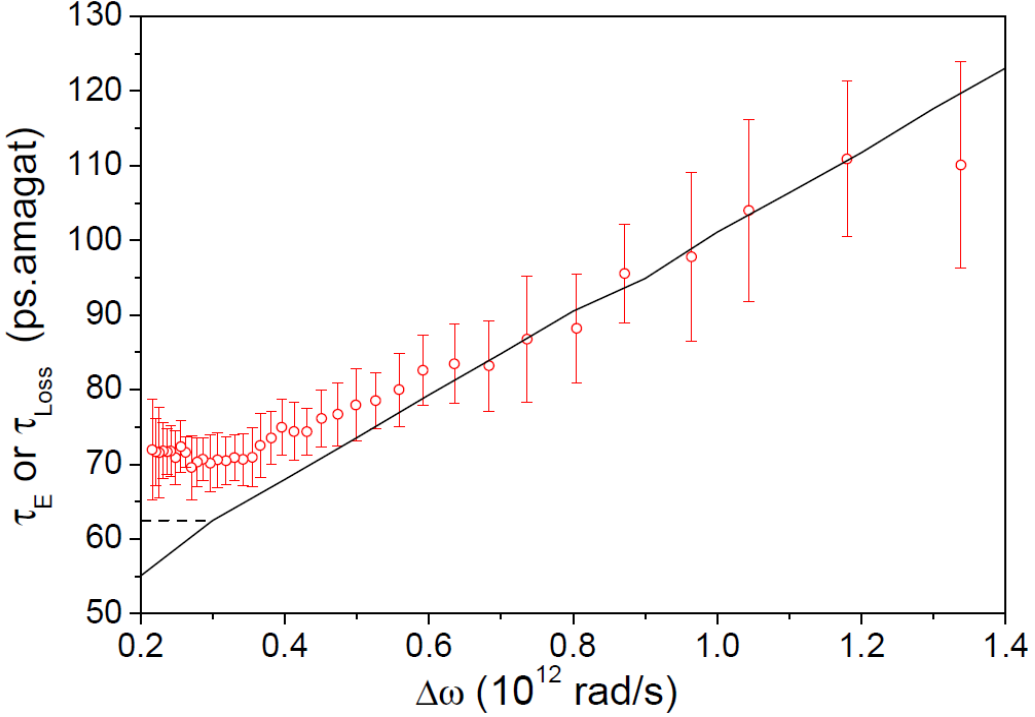

**Fig. 5:** Decay time constants of molecules' loss $\tau_{Loss}$ (black line) obtained from CMDS (see text) and of the measured (red open circles with error bars) echo decays from Fig. 2 versus the average filament width (see text).

Note that the experimentally observed plateau of $\tau_E$ for small values of $\Delta\omega$ in Fig. 5 (and for large delays in Fig. 2) is absent in the purely classical calculations of $\tau_{Loss}$ (and in the CMDS results of Fig. 2). This is because the CMDS allow filament widths to become smaller than the quantum limit $\hbar/I = 0.15 \ 10^{12}$ rad/s. However, *a posteriori* imposing this limit (i.e. setting $\tau_{Loss}(\Delta\omega) = \tau_{Loss}(\Delta\omega_L)$ for $\Delta\omega \leq \Delta\omega_L = 2\hbar/I$) creates a plateau below 0.3 $10^{12}$ rad/s in the black curve in Fig. 5 (shown by the dashed curve) making it more consistent with the measurements.

### V.B Classical-quantum correspondences

It is of interest to relate the present results based on a classical model to those obtained with a quantum approach [22] since, despite the fundamental difference between them, both well predict the observed behavior of the echoes as shown by Fig. 2 above and Fig. 3 of [22], respectively. This connection can be made, as detailed below, by considering that the $\langle J, M | \rho(t) | J' = J \pm 2, M \rangle$ quantum coherence involved in the alignment factor, with $\rho$ being the density operator, is represented by those molecules that classically rotate with angular speeds $\omega$ such that $2\omega$ is the closest to $|E_J - E_{J'}|/\hbar$ (the factor 2 being introduced since we here study the behavior of $<\cos^2\theta(t)>$). Now, the quantum results show that exchanges between coherences limit the echo decay, a nonsecular effect significant at early times which vanishes with increasing delay and the associated dephasing [22]. The classical equivalent of this is the fact that exchanges between rotational-speed classes limit the echo decay as long as they occur within a given filament, an event all the less probable when the filament width has narrowed with the elapsed time.

In order to demonstrate these statements, let us, for simplicity, consider the 2D case where a linear molecule classically rotates in a plane and denote by $\theta_m(t)$ and $\omega_m(t)$ the axis orientation and angular speed of molecule *m* at time *t*. Assuming that the molecules have been instantaneously aligned along the $\theta = 0$ direction by a laser pulse at *t*=0, one has, under collision-free conditions:

$$\theta_m(t>0) = \omega_m(t=0) \times t \tag{1}$$

and the associated contribution to the alignment $<\cos^2\theta(t)>$ factor thus oscillates with the angular frequency $2\omega_m(t=0)$. In the quantum world, Eq. (A12) of the Appendix shows that the contribution of the $\langle JM|\boldsymbol{\rho}(t)|J'M\rangle$ coherence to the alignment factor oscillates with the angular frequency $B[J(J+1) - J'(J'+1)]/\hbar = (E_J - E_{J'})/\hbar \equiv \omega_{J,J'}$. One may thus consider that a specific quantum coherence $\langle J_0 M|\boldsymbol{\rho}(t)|J_1 M\rangle$ is represented by those molecules that classically rotate with angular speed $\omega_m$ such that $2\omega_m$ is closer to $\omega_{J_0,J_1}$ than to $\omega_{J,J'}$ for any other $\langle JM|\boldsymbol{\rho}(t)|J'M\rangle$ coherence. The dephasing $(\omega_{J_0,J_1} - \omega_{J'_0,J'_1})t$ between to different quantum coherences then corresponds to twice the angle between the axes of the associated molecules. For rotors in 3D, an additional criterion concerning the orientation of the angular momentum is that it is constrained by the fact that the projection $\vec{\omega}_m \cdot \vec{Z}$ of $\vec{\omega}_m$ onto the space fixed $\vec{Z}$ axis should be such that $\vec{\omega}_m \cdot \vec{Z}/\|\vec{\omega}_m\|$ is the closest to $M/\bar{J}$. In this last term, since only the coherences with $J_1 = J_0 \pm 2$ contribute to the alignment factor, and since many levels are populated, the choice $\bar{J} = (J_0 + J_1)/2$ is reasonable.

Now that we have defined correspondences between quantum coherences and classically rotating molecules, let us consider the case where intermolecular collisions participate to the evolution of the system in order to relate nonsecular quantum effects to the widths of the filaments in the classical rotational phase space. Let us first note that, for the considered $N_2O$-He interacting pairs, the typical duration of a collision is less than 0.1 ps (as discussed in the Supplementary Material of Ref. [22]), much smaller that the rotational period of the molecules for the significantly populated levels. One may thus assume, consistently with the validity of the Infinite Order Sudden approximation used in Ref. [22], that collisions and the associated exchanges between coherences as well as those of the classical rotational motion are instantaneous.

Let us first consider a classically rotating molecule *m* whose angular speed is instantaneously changed from $\omega_m$ to $\omega'_m$ by a collision occurring at a given time $t_c$. Assuming that it had been rotating freely until $t_c$, so that Eq. (1) can be used until this time, and within a 2D case for simplicity, the axis orientations of this molecules for $t>t_c$ is then given by:

$$\theta_m(t>t_c) = [\omega_m(t=0) - \omega'_m]t_c + \omega'_m \times t \ . \tag{2}$$

With respect to those molecules $m'$, such that $\omega_{m'}(t=0) = \omega'_m$, which have been rotating freely up to *t*, the axis orientation difference is thus

$$\Delta\theta_{m,m'}(t_c) = [\omega_m(t=0) - \omega_{m'}(t=0)]t_c \ . \tag{3}$$

For the cumulative effect of such collision-induced changes not to "destroy" the echo by more than a given amount, $|\Delta\theta_{m,m'}(t_c)|$ should be smaller than a given threshold value $\Delta\theta_0$ discussed below. This implies that:

$$|\omega_m(t=0) - \omega_{m'}(t=0)| < \Delta\theta_0/t_c \ . \tag{4}$$

This enlightens the role played by the filaments, since echo-preserving collisions should keep the rotational speed change smaller than a given amount that decreases inversely proportionally with time, consistently with the CMDS results in Fig. 4(c). Precisely defining the value of $\Delta\theta_0$ is not possible, but one may consider that a choice such that the dephasing of the squared cosine should be less than π/2 is reasonable. This leads to $\Delta\theta_0 = \pi/4$, so that Eq. (4) becomes

$$|\omega_m(t=0) - \omega_{m'}(t=0)| < \sim 0.8/t_c \ , \tag{5}$$

which is consistent with the CMDS results in Fig. 4(c).

Let us now turn to the quantum coherences and the breakdown of the secular approximation at short times evidenced in [22]. Recall that the latter neglects all collision-induced exchanges between coherences (i.e. between nondiagonal elements of the density matrix **ρ**) as well as those between coherences and populations (diagonal elements of **ρ**), an approximation that becomes valid when $|\omega_{J_0,J'_0} - \omega_{J_1,J'_1}|t$ becomes greater than a few times $\pi$ [22]. In order to relate nonsecular effects to changes of the classical rotational speed, consider an instantaneous transfer, at time $t_c$, from the $(J_0, J'_0)$ to the $(J_1, J'_1)$ coherence. The latter, which has evolved freely until this time according to Eq. (A9) of the Appendix then becomes, for $t > t_c$:

$$\langle J_1 M | \rho(t > t_c) | J'_1 M \rangle = \langle J_1 M | \tilde{\rho}(t = 0^+) | J'_1 M \rangle e^{i\omega_{J_1,J'_1}t} \\ + F \langle J_0 M | \tilde{\rho}(t = 0^+) | J'_0 M \rangle e^{i(\omega_{J_0,J'_0} - \omega_{J_1,J'_1})t_c} e^{i\omega_{J_1,J'_1}t}, \quad (6)$$

where the first term corresponds to the free evolution of the $(J_1, J'_1)$ coherence and the second is the contribution coming from the $(J_0, J'_0)$ coherence (assuming that a fraction $F$ of its amplitude has been transferred). For the cumulative effect of such collision-induced transfers to limit the echo decay (with respect to the secular case where only losses are taken into account), $|\omega_{J_0,J'_0} - \omega_{J_1,J'_1}|t_c$ should be smaller than a given threshold value, which is the quantum equivalent of the collision-induced classical rotational speed change limit discussed above. In other words, the quantum secular approximation validity region is connected to the filament widths in the classical phase space. Nonsecular effects in the quantum world are significant at times such that the dephasing between coherences is small and they tend to become negligible for large dephasings. In the classical world, the amount of nonsecularity is driven by the filaments' widths, with the secular approximation becoming increasingly valid as the former becomes smaller and smaller.

## VI. CONCLUSION

We have shown, with very good agreement between theoretical predictions and measurements, that alignment echoes created in a molecular gas by two successive intense and short laser excitations enable to probe the ability of collisions to change the rotational speed by various amounts. This is thanks to the progressive narrowing of the filaments created, in the classical rotational phase space, by the first aligning pulse. With respect to the alignment revivals in the time domain and to the pressure-broadening of absorption lines in the spectral domain, echoes thus give a much more contrasted picture by enabling to track the dissipation at the very early stage of the collisional relaxation process. This unprecedented scrutiny opens renewed perspectives for our understanding of collisional processes and stringent tests of rotational-relaxation models and intermolecular potentials. Rotational alignment echoes being a generic classical phenomenon, this statement is also valid for very complex (heavy, nonlinear) molecules for which quantum calculations are intractable. This further broadens the potential applications of this study. In addition, it would be interesting to extend the present work to the dissipation of the alignment of molecules embedded inside helium nanodroplets, a topic of growing interest for which experimental results have been obtained recently (see Refs. [30-32] and references there-in). However, note that modeling the helium cage and the embedded molecule behavior at the involved very low temperature with a classical approach may be a challenging theoretical problem. The present paper also shows that the rotational alignment echoes are a bridge between the classical and quantum worlds, involving classical equivalents of quantum coherences. This property could be used for the theoretical modeling and understanding of nonsecular, but also of non Markovian, effects in systems for which quantum models are unavoidable or too complex.

**Acknowledgements:** This work was supported by the Conseil Régional de Bourgogne, the ERDF Operational Programme-Burgundy 2014/2020, and the EIPHI Graduate School (Contract No. "ANR-17-EURE-0002"). J. M. acknowledges support from the China Scholarship Council, and J. W. from the National Key R&D Program (Grant No. 2018YFA0306303) and the NSFC (Grants No. 11425416, No. 11761141004 and No. 11834004). J.-M. H. and T. D. benefited from the IPSL mésocentre ESPRI (Ensemble de Services Pour la Recherche à l'IPSL) facility for computer simulations.

# Appendix: On the nature of the revival and echo phenomena

The time evolution of the density matrix $\boldsymbol{\rho}(t)$ of linear molecules subject to a nonresonant and linearly polarized laser pulse is, under collision-free conditions, given by:

$$\frac{d\boldsymbol{\rho}}{dt}(t) = -\frac{i}{\hbar}[\mathbf{H}_0 + \mathbf{H}_L(t), \boldsymbol{\rho}(t)] , \qquad (A1)$$

where $\mathbf{H}_0$ is the free rotation Hamiltonian and:

$$\mathbf{H}_L(t) = -\frac{1}{4}\Delta\alpha E^2(t)(\cos^2\theta - 1/3) = -\frac{1}{6}\Delta\alpha E^2(t)\mathbf{P}_2(\cos\theta) , \qquad (A2)$$

with $\Delta\alpha$ the anisotropic polarizability, $E(t)$ the envelop of the laser pulse and $\mathbf{P}_2$ a Legendre polynomial. Writing Eq. (A1) in the interaction picture, introducing

$$\tilde{\boldsymbol{\rho}}(t) = e^{+i\mathbf{H}_0 t/\hbar}\boldsymbol{\rho}(t)e^{-i\mathbf{H}_0 t/\hbar} , \qquad (A3)$$

and assuming a laser pulse applied at $t=0$ with a negligible duration (sudden approximation) leads, just after the pulse, to:

$$\tilde{\boldsymbol{\rho}}(t=0^+) = e^{+iC\mathbf{P}_2(\cos\theta)}\tilde{\boldsymbol{\rho}}(t=0^-)e^{-iC\mathbf{P}_2(\cos\theta)} , \qquad (A4)$$

where $C$ is proportional to $\Delta\alpha$ and to the energy of the pulse, $\tilde{\boldsymbol{\rho}}(t=0^-) = e^{-\mathbf{H}_0/k_B T}/\mathrm{Tr}(e^{-\mathbf{H}_0/k_B T})$ where $\mathrm{Tr}(...)$ denotes the trace, $k_B$ being the Boltzmann constant and $T$ the temperature. Recall that, since

$$\langle JM|\mathbf{P}_2(\cos\theta)|J'M'\rangle = \delta_{M,M'}(-1)^M\sqrt{(2J+1)(2J'+1)}\begin{pmatrix}J' & 2 & J\\ 0 & 0 & 0\end{pmatrix}\begin{pmatrix}J' & 2 & J\\ M' & 0 & -M\end{pmatrix} , \qquad (A5)$$

where $(:::)$ is a 3J symbol and $J$ and $M$ are the quantum numbers associated with the rotational angular momentum and with its projection along the quantification axis, one has the selection rules $M' = M$ and $J' = J, J\pm 2$. We now make a development of the sudden approximation operators, i.e.:

$$e^{\pm iC\mathbf{P}_2(\cos\theta)} = 1 \pm iC\mathbf{P}_2(\cos\theta) - \frac{1}{2}C^2\mathbf{P}_2(\cos\theta)^2 + ... , \qquad (A6)$$

and limit ourselves to the first order. The matrix elements of $\boldsymbol{\rho}(t)$ defined in Eq. (A4) then become:

$$\langle JM|\tilde{\boldsymbol{\rho}}(t=0^+)|J'M\rangle = \delta_{J,J'}\langle JM|\tilde{\boldsymbol{\rho}}(t=0^-)|JM\rangle$$
$$+ iC\langle JM|\mathbf{P}_2(\cos\theta)|J'M\rangle\left[\langle J'M|\tilde{\boldsymbol{\rho}}(t=0^-)|J'M\rangle - \langle JM|\tilde{\boldsymbol{\rho}}(t=0^-)|JM\rangle\right] . \qquad (A7)$$

After the pulse, the system evolves freely and one has:

$$\boldsymbol{\rho}(t>0^+) = e^{-i\mathbf{H}_0 t/\hbar}\tilde{\boldsymbol{\rho}}(t=0^+)e^{i\mathbf{H}_0 t/\hbar} , \qquad (A8)$$

i.e. for a rigid rotor of rotational constant $B$:

$$\langle JM|\boldsymbol{\rho}(t>0^+)|J'M\rangle = \langle JM|\tilde{\boldsymbol{\rho}}(t=0^+)|J'M\rangle e^{iBt[J'(J'+1)-J(J+1)]/\hbar} . \qquad (A9)$$

## 1. Single pulse and the alignment revivals

When a single excitation pulse is applied, Eq. (A9) leads to the following of alignment factor:

$$\langle\cos^2\theta\rangle(t>0^+) = \sum_{J,M}\langle JM|\boldsymbol{\rho}(t)\cos^2\theta|JM\rangle = \sum_{J,M,J'}\langle JM|\boldsymbol{\rho}(t)|J'M\rangle\langle J'M|\cos^2\theta|JM\rangle$$
$$= \sum_{J,M,J'}\langle JM|\tilde{\boldsymbol{\rho}}(t=0^+)|J'M\rangle e^{iBt[J'(J'+1)-J(J+1)]/\hbar}\langle J'M|\cos^2\theta|JM\rangle , \qquad (A10)$$

in which the 1/3 contribution coming from the zeroth-order terms in Eqs. (A6) and (A7) has been disregarded. Introducing Eq. (A7) into Eq. (A10) only keeping the terms proportional to the laser energy (i.e. to C) leads to:

$$\langle\cos^2\theta\rangle(t>0^+) = +iC \sum_{J,M,J'} \langle JM|\mathbf{P}_2(\cos\theta)|J'M\rangle\langle J'M|\tilde{\boldsymbol{\rho}}(t=0^-)|J'M\rangle e^{iBt[J'(J'+1)-J(J+1)]/\hbar}\langle J'M|\cos^2\theta|JM\rangle$$
$$-iC \sum_{J,M,J'} \langle JM|\mathbf{P}_2(\cos\theta)|J'M\rangle\langle JM|\tilde{\boldsymbol{\rho}}(t=0^-)|JM\rangle e^{iBt[J'(J'+1)-J(J+1)]/\hbar}\langle J'M|\cos^2\theta|JM\rangle$$
(A11)

By permuting $J$ and $J'$ in the second sum taking into account that $\langle JM|\mathbf{P}_2(\cos\theta)|J'M\rangle = \langle J'M|\mathbf{P}_2(\cos\theta)|JM\rangle$ and $\langle JM|\cos^2\theta|J'M\rangle = \langle J'M|\cos^2\theta|JM\rangle$, Eq. (A11) reduces to:

$$\langle\cos^2\theta\rangle(t>0^+) = 2C \sum_{J,M,J'=J\pm 2} \langle JM|\mathbf{P}_2(\cos\theta)|J'M\rangle\langle JM|\tilde{\boldsymbol{\rho}}(t=0^-)|JM\rangle$$
$$\times\langle J'M|\cos^2\theta|JM\rangle \sin\{Bt[J(J+1)-J'(J'+1)]/\hbar\}$$
(A12)

It is obvious that all the oscillating terms of the sum in Eq. (A12) rephase, all sine functions being zero, for times $t$ multiples of $t_{\text{rev}} = h/(4B)$ for $N_2O$ [$t_{\text{rev}} = h/(8B)$ in the case of $CO_2$ for which only even $J$ values exist]. This leads, with $B(N_2O)=0.42$ cm$^{-1}$ and $B(CO_2)=0.39$ cm$^{-1}$, to the so-called "revivals" with (positive and negative) extrema in the alignment factor appearing around multiples of $t_{\text{rev}} \approx 20$ ps [22,33] and $t_{\text{rev}} \approx 11$ ps [8,15] ps, respectively. Since these specific times depend on the rotational constant $B$, the revivals are of purely quantum nature. Equation (A12) shows that the amplitudes of the revivals are, in the weak field limit, proportional to the laser energy [through the multiplicative factor $C$], in agreement with measurements [34] and direct calculations [15] of the alignment factor. Also note that the time-independent component of the alignment can be obtained using the same approach, but results from the second-order terms. The resulting "permanent"-alignment amplitude is thus proportional to the square of the laser energy, a finding also consistent with experiments [34] and independent calculations [15].

**2. Two pulses and the echoes**

First note that Eq. (A4) can be rewritten as:
$$\tilde{\boldsymbol{\rho}}(t) = \mathbf{U}\tilde{\boldsymbol{\rho}}(t=0^-)\mathbf{U}^*,$$
(A13)

with $\mathbf{U} = e^{+iC\mathbf{P}_2(\cos\theta)}$. Let us now consider the case where a second pulse is applied at $t=\tau_{12}$. From Eqs. (A8), (A10), and (A13), one obtains:

$$\boldsymbol{\rho}(t>\tau_{12}) = e^{-i\mathbf{H}_0(t-\tau_{12})/\hbar}\mathbf{U}_2 e^{-i\mathbf{H}_0\tau_{12}/\hbar}\mathbf{U}_1\tilde{\boldsymbol{\rho}}(t=0^-)\mathbf{U}_1^* e^{+i\mathbf{H}_0\tau_{12}/\hbar}\mathbf{U}_2^* e^{+i\mathbf{H}_0(t-\tau_{12})/\hbar},$$
(A14)

and:
$$\langle\cos^2\theta\rangle(t>\tau_{12}) = \sum_{J,J',J'',J''',J'''',M} A_{J,J'',J''',J'''',J'}\langle J'M|\cos^2\theta|JM\rangle$$
$$\times\exp\{-iB[J(J+1)(t-\tau_{12})+J''(J''+1)\tau_{12}-J''''(J''''+1)\tau_{12}-J'(J'+1)(t-\tau_{12})]/\hbar\}$$
(A15)

with
$$A_{J,J'',J''',J'''',J'} = \langle JM|\mathbf{U}_2|J''M\rangle\langle J''M|\mathbf{U}_1|J'''M\rangle\langle J'''M|\tilde{\boldsymbol{\rho}}(t=0^-)|J'''M\rangle$$
$$\times\langle J'''M|\mathbf{U}_1^*|J''''M\rangle\langle J''''M|\mathbf{U}_2^*|J'M\rangle$$
(A16)

Let us ne now use
$$\langle JM|\mathbf{U}_n|J'M\rangle = \delta_{J,J'} + iC_n\langle JM|\mathbf{P}_2(\cos\theta)|J'M\rangle - C_n^2\langle JM|\mathbf{P}_2(\cos\theta)^2|J'M\rangle/2,$$
(A17)

where $C_n$ is proportional to the intensity of the n$^{\text{th}}$ pulse, and consider the various orders in $C_1$ and $C_2$. The preceding section shows that the contributions to Eq. (A15) which are proportional to $C_1$ and $C_1^2$ (respectively to $C_2$ and $C_2^2$) lead to the revivals and to the permanent alignment generated by the first (respectively second) pulse, respectively. Considering the four terms proportional to $C_1C_2$, it is "relatively" easy to show that they cancel out. Let us now look at the contributions proportional to $C_1C_2^2$, coming from approximating $\mathbf{U}_1$ or $\mathbf{U}_1^*$ by $\pm iC_1\mathbf{P}_2(\cos\theta)$. There are six of them, two that result from approximating both $\mathbf{U}_2$ and $\mathbf{U}_2^*$ by $\pm iC_2\mathbf{P}_2(\cos\theta)$, and four others obtained when either $\mathbf{U}_2$ or $\mathbf{U}_2^*$ is approximated by $-C_2^2 \mathbf{P}_2(\cos\theta)^2/2$ while $\mathbf{U}_2^*$ or $\mathbf{U}_2$ is approximated by unity.

Let us first consider the first two, which are given by Eqs. (A15) and (A16) with either $J'' = J'''$ or $J''' = J''''$ since only the first order contributions in $C_1$ are kept, i.e.:

$$\langle \cos^2\theta \rangle (t > \tau_{12}) = \sum_{J,J',J'',J''',M} A_{J,J'',J'',J''',J'} \langle J'M | \cos^2\theta | JM \rangle$$
$$\times \exp\{-iB[J(J+1)(t-\tau_{12}) + J''(J''+1)\tau_{12} - J''''(J''''+1)\tau_{12} - J'(J'+1)(t-\tau_{12})]/\hbar\}$$
$$+ \sum_{J,J',J'',J''',M} A_{J,J'',J''',J''',J'} \langle J'M | \cos^2\theta | JM \rangle \qquad (A18)$$
$$\times \exp\{-iB[J(J+1)(t-\tau_{12}) + J''(J''+1)\tau_{12} - J''''(J''''+1)\tau_{12} - J'(J'+1)(t-\tau_{12})]/\hbar\}$$

with

$$A_{J,J'',J'',J''',J'} = -iC_1C_2^2 \langle JM | \mathbf{P}_2(\cos\theta) | J''M \rangle \langle J''M | \tilde{\boldsymbol{\rho}}(t=0^-) | J''M \rangle$$
$$\times \langle J''M | \mathbf{P}_2(\cos\theta) | J''''M \rangle \langle J''''M | \mathbf{P}_2(\cos\theta) | J'M \rangle \qquad (A19)$$
$$A_{J,J'',J''',J''',J'} = iC_1C_2^2 \langle JM | \mathbf{P}_2(\cos\theta) | J''M \rangle \langle J''M | \mathbf{P}_2(\cos\theta) | J''''M \rangle$$
$$\times \langle J''''M | \tilde{\boldsymbol{\rho}}(t=0^-) | J''''M \rangle \langle J''''M | \mathbf{P}_2(\cos\theta) | J'M \rangle$$

By making the changes $J \leftrightarrow J'$ and $J'' \leftrightarrow J''''$ in the second sum in Eq. (A18), one obtains:

$$\langle \cos^2\theta \rangle (t > \tau_{12}) = 2C_1C_2^2 \sum_{J,J',J'',J''',M} \langle J''M | \tilde{\boldsymbol{\rho}}(t=0^-) | J''M \rangle \langle J''M | \mathbf{P}_2(\cos\theta) | J''''M \rangle \langle J''''M | \mathbf{P}_2(\cos\theta) | J'M \rangle$$
$$\times \langle JM | \mathbf{P}_2(\cos\theta) | J''M \rangle \langle J'M | \cos^2\theta | JM \rangle \sin\{-B[J(J+1)(t-\tau_{12}) + J''(J''+1)\tau_{12} - J''''(J''''+1)\tau_{12} - J'(J'+1)(t-\tau_{12})]/\hbar\} \qquad (A20)$$

It is easy to show that, for the *specific* time $t = 2\tau_{12}$, the phase of the sine function is exactly zero, *regardless of* the values of $J$, $J'$ and $B$, for the terms associated with $J'' = J'$ and $J'''' = J$ (e.g. $J' = J'' = J \pm 2 = J'''' \pm 2$ that are allowed by the selection rule of the various $\mathbf{P}_2(\cos\theta)$ matrix elements involved). Note that $J = J'$ and $J'''' = J''$ is also allowed but that this choice leads to a time-independent ("permanent") alignment. Close to $t = 2\tau_{12}$, at $t = 2\tau_{12} \pm \Delta t$ the associated alignment factor shows two antisymmetric extrema that are the echo signal (e.g. Fig. 1). The fact that its central position does not depend on $B$ (as well as on the temperature and laser pulses intensities), shows that its existence is of *classical* nature. This explains why the echoes are predicted by purely classical calculations (e.g. right hand-side panel of Fig. 3 and Refs. [14,18]). However, the fact that both $\Delta t$ and the amplitudes of the extrema depend on $B$ make the echo "semi-classical".

Let us now consider the other terms, associated with $J'' = J'''$ or $J''' = J''''$ while $J'' = J$ or $J' = J''''$. At the time $t = 2\tau_{12}$ of the echo, they are given by:

$$\langle \cos^2\theta \rangle(2\tau_{12}) = -(iC_1C_2^2/2) \sum_{J,J',J'',M} \langle J'M|\cos^2\theta|JM\rangle$$

$$\times \Big[ +\langle JM|\mathbf{P}_2^2(\cos\theta)|J''M\rangle\langle J''M|\mathbf{P}_2(\cos\theta)|J'M\rangle\langle J'M|\tilde{\boldsymbol{\rho}}(t=0^-)|J'M\rangle$$

$$\times \exp\{-iB\tau_{12}[J(J+1)+J''(J''+1)-2J'(J'+1)]/\hbar\}$$

$$-\langle JM|\tilde{\boldsymbol{\rho}}(t=0^-)|JM\rangle\langle JM|\mathbf{P}_2(\cos\theta)|J''M\rangle\langle J''M|\mathbf{P}_2^2(\cos\theta)|J'M\rangle$$

$$\times \exp\{-iB\tau_{12}[2J(J+1)-J''(J''+1)\tau_{12}-J'(J'+1)]/\hbar\} \qquad . \qquad (A21)$$

$$-\langle JM|\mathbf{P}_2^2(\cos\theta)|J''M\rangle\langle J''M|\tilde{\boldsymbol{\rho}}(t=0^-)|J''M\rangle\langle J''M|\mathbf{P}_2(\cos\theta)|J'M\rangle$$

$$\times \exp\{-iB\tau_{12}[J(J+1)+J''(J''+1)-2J'(J'+1)]/\hbar\}$$

$$+\langle JM|\mathbf{P}_2(\cos\theta)|J''M\rangle\langle J''M|\tilde{\boldsymbol{\rho}}(t=0^-)|J''M\rangle\langle J''M|\mathbf{P}_2^2(\cos\theta)|J'M\rangle$$

$$\times \exp\{-iB\tau_{12}[2J(J+1)-J'''(J'''+1)-J'(J'+1)]/\hbar\} \Big]$$

Making the change $J \leftrightarrow J'$ in the second and last terms enables to combine them with the first and third, respectively, leading to:

$$\langle \cos^2\theta \rangle(2\tau_{12}) = C_1C_2^2 \sum_{J,J',J'',M} \langle J'M|\cos^2\theta|JM\rangle$$

$$\times \Big[ +\langle JM|\mathbf{P}_2^2(\cos\theta)|J''M\rangle\langle J''M|\mathbf{P}_2(\cos\theta)|J'M\rangle\langle J'M|\tilde{\boldsymbol{\rho}}(t=0^-)|J'M\rangle$$

$$\times \sin\{-B\tau_{12}[J(J+1)+J''(J''+1)-2J'(J'+1)]/\hbar\} \qquad , \qquad (A22)$$

$$+\langle JM|\mathbf{P}_2(\cos\theta)|J''M\rangle\langle J''M|\tilde{\boldsymbol{\rho}}(t=0^-)|J''M\rangle\langle J''M|\mathbf{P}_2^2(\cos\theta)|J'M\rangle$$

$$\times \sin\{-B\tau_{12}[2J(J+1)-J'''(J'''+1)-J'(J'+1)]/\hbar\} \Big]$$

and, by again making the change $J \leftrightarrow J'$ in the second term:

$$\langle \cos^2\theta \rangle(2\tau_{12}) = C_1C_2^2 \sum_{J,J',J'',M} \langle J'M|\cos^2\theta|JM\rangle \times \sin\{-B\tau_{12}[J(J+1)+J''(J''+1)-2J'(J'+1)]/\hbar\}$$

$$\times \langle JM|\mathbf{P}_2^2(\cos\theta)|J''M\rangle\langle J'M|\mathbf{P}_2(\cos\theta)|J''M\rangle\Big[\langle J'M|\tilde{\boldsymbol{\rho}}(t=0^-)|J'M\rangle-\langle J''M|\tilde{\boldsymbol{\rho}}(t=0^-)|J''M\rangle\Big] \qquad . \qquad (A23)$$

Among the various terms, the only ones for which the phase of the sine function is independent of $J$ are those with $J'=J\pm 2$ and $J''=J\pm 4$. For these specific values of $J'$ and $J''$, the sine becomes $\sin(-8B\tau_{12}/\hbar)$. We see that these contributions to the echo at $t=2\tau_{12}$ have amplitudes which, contrary to the ones discussed above, depend on $\tau_{12}$. They have a periodicity of $h/(8B)$ with respect to $\tau_{12}$ and lead to a switching of the extrema, from minimum/maximum to maximum/minimum, on both sides of $t=2\tau_{12}$ for $\tau_{12}=h/(32B)$ and $\tau_{12}=3h/(32B)$. This contribution will thus be constructive or destructive, with respect to those discussed above, with maxima or minima at $\tau_{12}=h/(16B)$. Its direct dependence on the rotational constant $B$ shows that it is of quantum nature.

In conclusion, we have shown that two pulses at $t=0$ and $t=\tau_{12}$ generate an echo in the alignment factor at $t=2\tau_{12}$ regardless of the molecule rotational constant, but also of the laser pulses intensities and of the temperature. This independence, confirmed by experiments [13,14,22], shows that this echo is of *classical* nature. We also demonstrated that the echo amplitude is, in the weak field limit, proportional to the energy of the first pulse and to the square of the energy of the second pulse, a result in agreement with experimental results [13,35] and with a classical model [18]. Finally, we showed that some contributions to the echo, which have amplitudes that depend on the rotational constant, have a periodicity of $h/(8B)$ with respect to $\tau_{12}$ and lead to extrema of the echo amplitude for $\tau_{12}=(2k+1)h/(16B)$, two results confirmed by calculations and experiments [13].